\documentclass[a4paper]{article}
\usepackage{icrc2013}
\pdfoutput=1

\usepackage[english]{babel}

\title{The Tunka Radio Extension (Tunka-Rex): Status and First Results}

\newcommand{\etal}{\MakeLowercase{\textit{et al. }}} 
\shorttitle{F.G.~Schr\"oder \etal Tunka-Rex: Status and First Results}

\authors{F.G.~Schr\"oder$^{1}$, N.M.~Budnev$^{2}$, O.A.~Gress$^{2}$, A.~Haungs$^{1}$, R.~Hiller$^{1}$, T.~Huege$^{1}$, Y.~Kazarina$^{2}$, M.~Kleifges$^{3}$, A.~Konstantinov$^{4}$, E.N.~Konstantinov$^{2}$, E.E.~Korosteleva$^{4}$, D.~Kostunin$^{1}$, O.~Kr\"omer$^{3}$, L.A.~Kuzmichev$^{4}$, R.R.~Mirgazov$^{2}$, L.~Pankov$^{2}$, V.V.~Prosin$^{4}$, G.I.~Rubtsov$^{5}$, C.~R\"uhle$^{3}$, E.~Svetnitsky$^{2}$, R.~Wischnewski$^{6}$, A.~Zagorodnikov$^{2}$ (Tunka-Rex Collaboration)}

\afiliations{$^{1}$Institut f\"ur Kernphysik, Karlsruhe Institute of Technology (KIT), Germany\\ $^{2}$Institute of Applied Physics ISU, Irkutsk, Russia\\ $^{3}$Institut f\"ur Prozessdatenverarbeitung und Elektronik, Karlsruhe Institute of Technology (KIT), Germany\\ $^{4}$Skobeltsyn Institute of Nuclear Physics MSU, Moscow, Russia\\ $^{5}$Institute for Nuclear Research of the Russian Academy of Sciences, Moscow, Russia\\ $^{6}$DESY, Zeuthen, Germany}

\email{frank.schroeder@kit.edu}

\abstract{Tunka-Rex is a new radio antenna array which extends the Tunka experiment in Siberia close to lake Baikal. It consists of $20$ antennas on an area of $1\,$km\textsuperscript{2} which measure the radio emission of high-energy air showers. Tunka-Rex is triggered by the photomultiplier array of Tunka measuring air-Cherenkov light of air showers in the energy range from about $10^{16}\,$eV to $10^{18}\,$eV. This configuration allows for the worldwide first hybrid measurements of the radio and air-Cherenkov signal for the same events: an ideal situation to perform a cross-calibration between both methods. Consequently, the main goal of Tunka-Rex is to determine the achievable energy and $X_\mathrm{max}$ precision of radio measurements by comparing them to the reconstruction of the air-Cherenkov measurements. Tunka-Rex started operation in autumn 2012, and already detected air-shower events. In this paper we present the status of Tunka-Rex and first results which indicate that Tunka-Rex measures indeed the radio emission by air showers and that is is sensitive to their energy.}
\keywords{Tunka-Rex, Tunka, ultra-high energy cosmic rays, extensive air showers, radio detection}

\begin{document}
\maketitle

\section{Introduction}
One of the key questions for the usability of the radio technique for cosmic-ray physics at ultra-high energies is whether radio measurements can provide a sufficient reconstruction precision for the properties of the primary cosmic-ray particles (arrival direction, energy, mass). Other radio experiments, e.g., LOPES \cite{FalckeNature2005} at the Karlsruhe Institute of Technology in Germany, could already prove that digital antenna arrays can detect air-showers with an energy threshold of approximately $10^{17}\,$eV. The achieved precision for the direction is less than $1^\circ$, which already is more than sufficient for current questions in astroparticle physics of charged cosmic rays \cite{SchroederLOPES_ARENA2012}. Other experiments could also show that it is in principle possible to reconstruct the energy with radio measurements \cite{SchroederLOPES_ARENA2012, GlaserAERA_ARENA2012, RebaiCODALEMAenergy2012}, and $X_\mathrm{max}$, the atmospheric depth of the shower maximum  \cite{ApelLOPES_MTD2012, PalmieriLOPES_ICRC2013}, a quantity sensitive to the primary mass. However, they could only give upper limits for the precisions ($20\,\%$ for the energy and about $100\,$g/cm\textsuperscript{2} for $X_\mathrm{max}$). This would be worse than what is reached by other detection techniques, but at the moment only the upper limits are known. It is open which precision for the energy and the mass composition can be achieved by radio arrays in radio-quiet environments. In particular, it is an open question whether the precision of the radio technique can be competitive to other techniques like air-fluorescence and air-Cherenkov-light measurements, which in contrast to the radio measurements are only possible in dark moonless nights.

The main goal of Tunka-Rex, the radio extension of the Tunka observatories for air showers, is to answer this question, i.e., to determine the precision for the energy and the atmospheric depth of the shower maximum $X_\mathrm{max}$ for the Tunka-Rex measurements. For this purpose, Tunka-Rex is built at the location of the Tunka-133 photomultiplier array measuring the air-Cherenkov light of air showers in the energy range between $10^{16}$ and $10^{18}\,$eV \cite{TunkaICRC2013}. Data of both detectors are recorded by a shared data-acquisition system, and the radio antennas are triggered by the photomultiplier measurements. This setup automatically provides hybrid measurements of the radio and the air-Cherenkov signal, and consequently allows a cross-calibration of both techniques. In particular, we can test the sensitivity of the Tunka-Rex radio measurements for the energy and for $X_\mathrm{max}$ by comparing them to the measurements of the established air-Cherenkov array.

\begin{figure*}[t]
\centering
\includegraphics[width=0.65\linewidth]{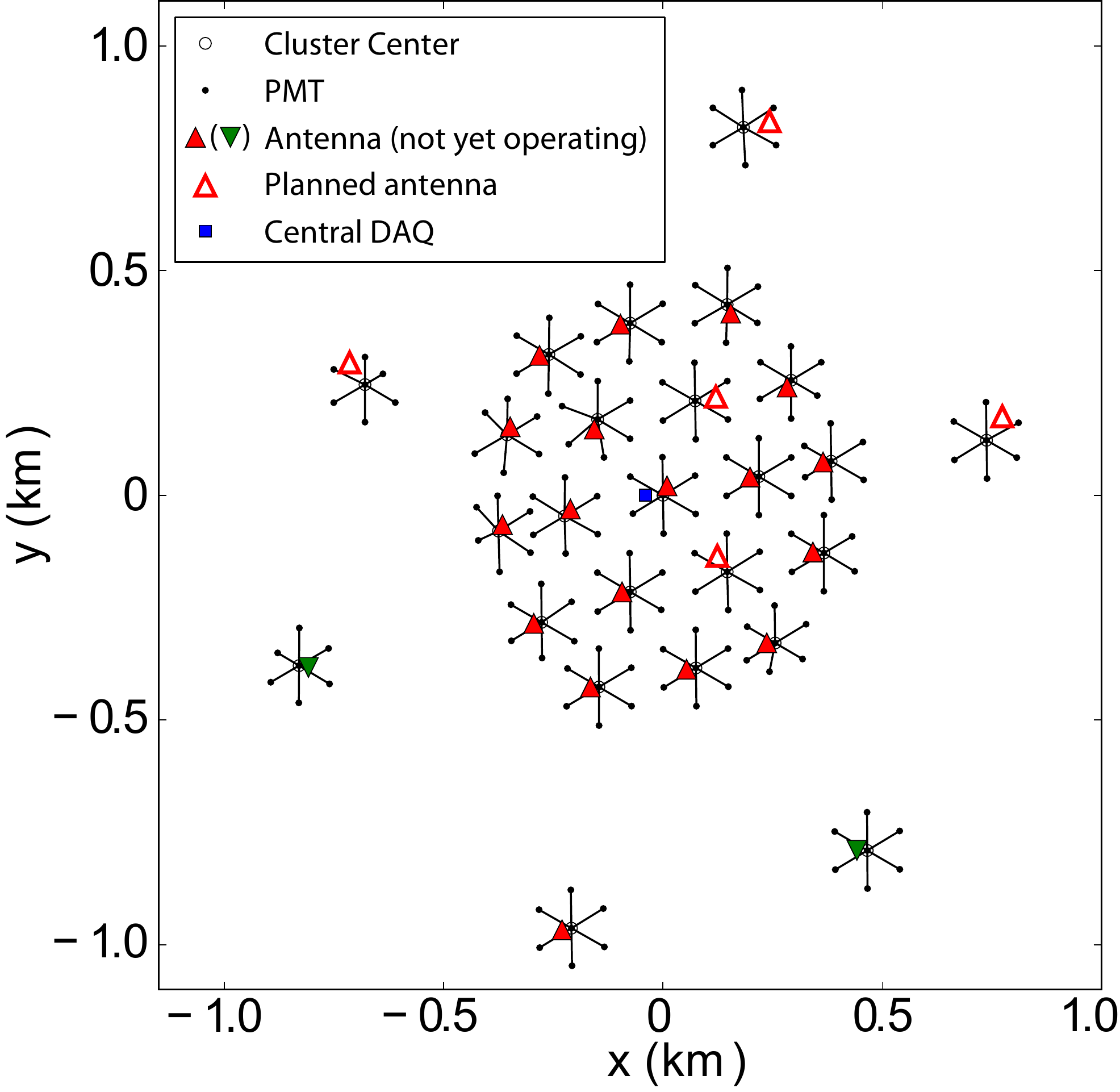}
\caption{Map of Tunka. One Tunka-Rex antenna, consisting of two SALLAs, is attached to each cluster of seven non-imaging photo-multipliers (PMTs). In the season from October 2012 to April 2013, 18 out of 20 Tunka-Rex antennas have been operating. In autumn 2013, five additional antennas will be installed to complete the array.} \label{fig_map}
\end{figure*}

\section{Setup and Status}
Tunka-Rex currently consists of 20 antennas attached by cables to the data-acquisition system (DAQ) of the Tunka photomultiplier (PMT) array (Fig.~\ref{fig_map}), which is organized in 25 clusters formed by 7 PMT each. The spacing between the antennas in the inner clusters is approximately $200\,$m, covering an area of roughly $1\,$km\textsuperscript{2}. At each antenna position there are two orthogonally aligned SALLAs \cite{AERAantennaPaper2012}, which enables a reconstruction of the electric field vector provided that the arrival direction of the signal is known. The SALLA has been chosen as antenna not only for economic reasons, but also because its properties depend only little on environmental conditions. Tunka-Rex is triggered by the photomultipliers and records the radio signal from the air showers between $30\,$ and $80\,$MHz, where the signal outside of this band is suppressed with an analog filter. First, this improves the signal-to-noise ratio of the air-shower radio signals, second, this ensures that we measure in the first Nyquist domain and can fully reconstruct the signal in this frequency band. Each of the clusters features its own local DAQ. There the signal from both, the antennas and the PMTs, is digitized and transmitted to the central DAQ where it is stored on disk. For more details on the Tunka-Rex hardware and how it impacts the systematic measurement uncertainties, see Ref.~\cite{HillerTunkaRex_ICRC2013}.

\begin{table}
\centering
\caption{Statistics of Tunka-Rex events passing the quality cuts in dependence of the zenith angle $\theta$, excluding the period from 08 to 24 Oct 2012 used for commissioning of the detector. The effective measurement time is limited by the PMT array, i.e.~light (moon) and weather conditions.} \label{tab_EventStatistics}
\vspace{0.1 cm}
\begin{tabular}{lccc} 
       &effective&\multicolumn{2}{c}{number of events}\\
measurement period &time& $\theta \le 50^\circ$ & $\theta > 50^\circ$\\
\hline
06 - 23 Nov 2012 & 56 h & 9 & 11 \\
04 - 23 Dec 2012 & 65 h & 8 & 12 \\
03 - 21 Jan 2013 & 114 h & 14 & 23 \\
01 - 17 Feb 2013 & 87 h & 12 & 22 \\
01 - 17 Mar 2013 & 70 h & 6 & 14 \\
\hline
Total sum & 392 h & 49 & 82 \\
\hline
\end{tabular}
\end{table}

\begin{figure*}[pht!]
\centering
\includegraphics[width=\linewidth]{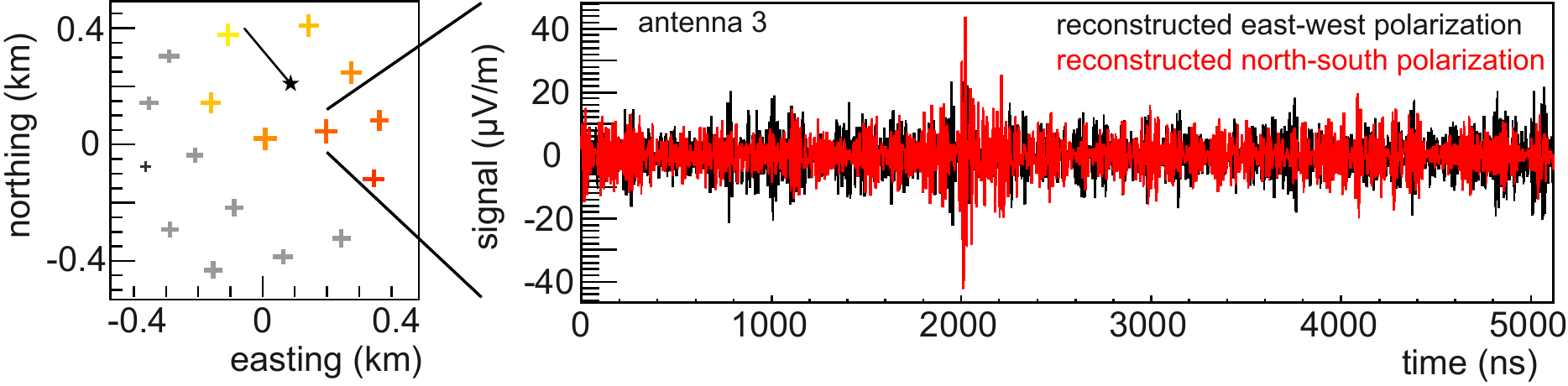}
\caption{Example for a Tunka-Rex event. Left: Footprint of the event, where the size of the crosses indicates the signal strength, the color code the arrival time, and the line the direction and shower core. Right: trace of the electric-field strength measured at the antenna with the strongest signal (at about 2000 ns).} \label{fig_exampleEvent}
\end{figure*}

Tunka-Rex started operation on 8 October 2012. Since then operates whenever the PMT array is operating, i.e. in dark moonless nights with good weather excluding the summer months from May to September. Only a small fraction of the air-Cherenkov events have also a clear radio signal (see Fig.~\ref{fig_exampleEvent} for an example). The reason is that the amplitude of the radio signal depends on the energy and the geometry of the air shower, in particular, the angle between the shower axis and the geomagnetic field, the shower inclination, and the relative position of the antennas. For a first analysis we used only high quality events which have a clear radio signal (amplitude / background $> 4$) in at least three antennas. Moreover we demand that the direction reconstructed from the arrival times of the radio signal agrees within $5^\circ$ with the direction obtained from the photomultiplier array. This cut excludes most of the background events, which by chance pass the signal-to-noise cut. In future, we plan to develop improved quality criteria based on the radio signal alone, to distinguish real from false events. For analysis of the radio measurements, we use the radio extension of the Offline software framework developed by the Pierre Auger Collaboration \cite{AugerOffline2007, RadioOffline2011}. It features a correction of the measured radio signal for the properties of the used hardware and a reconstruction of the electric field-strength vector at each antenna position.

\begin{figure}[p!]
\centering
\includegraphics[width=\linewidth]{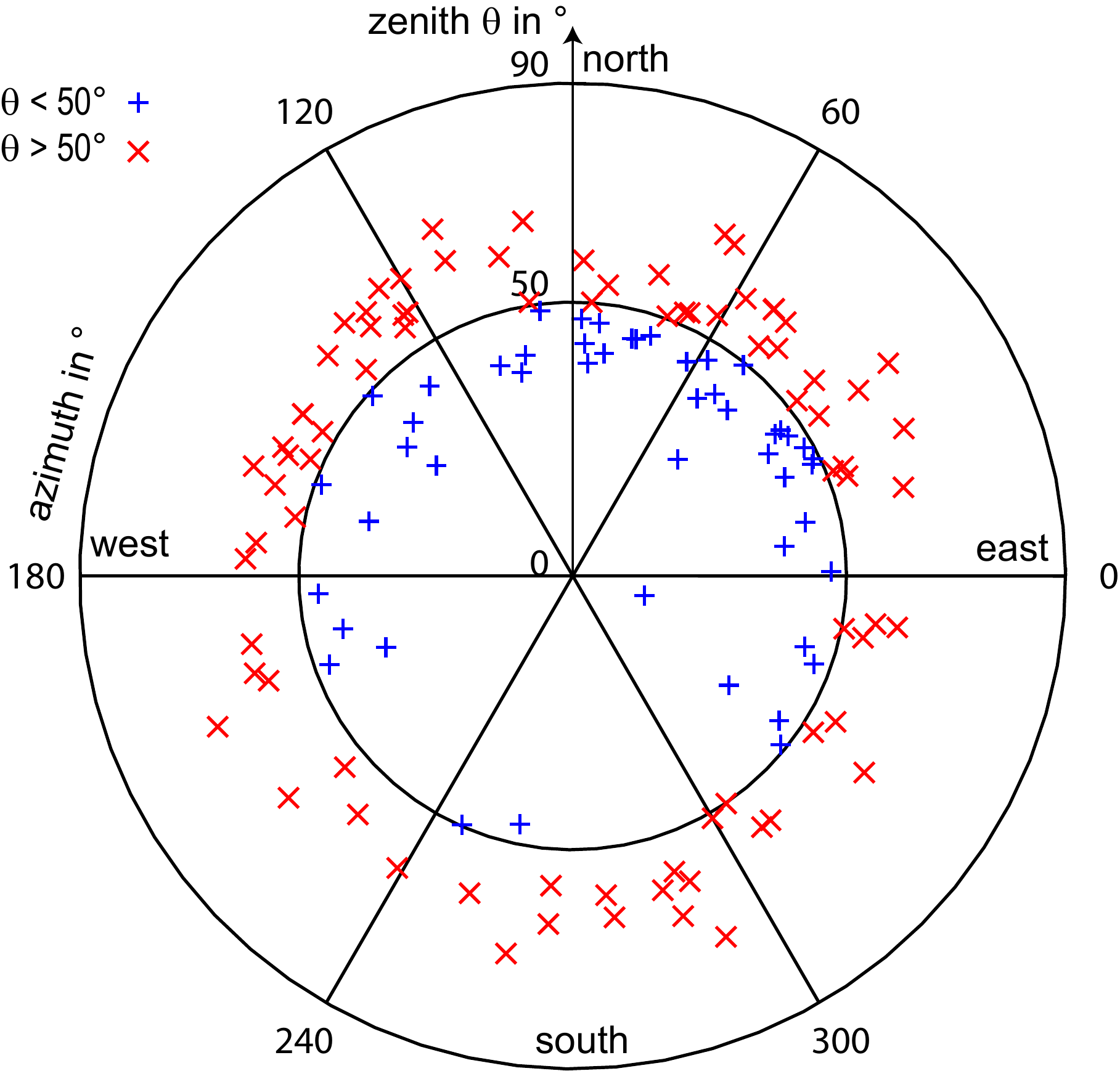}
\caption{Arrival directions of the Tunka-Rex events passing the quality cuts. Due to the geomagnetic effect, the radio signal is expected to be on average stronger for events coming from North, which explains the asymmetry in the detection efficiency: 89 of the 131 events are in the northern half of the plot, and 42 in the southern half.} \label{fig_angularDistribution}
\end{figure}

\begin{figure}[p!]
\centering
\includegraphics[width=\linewidth]{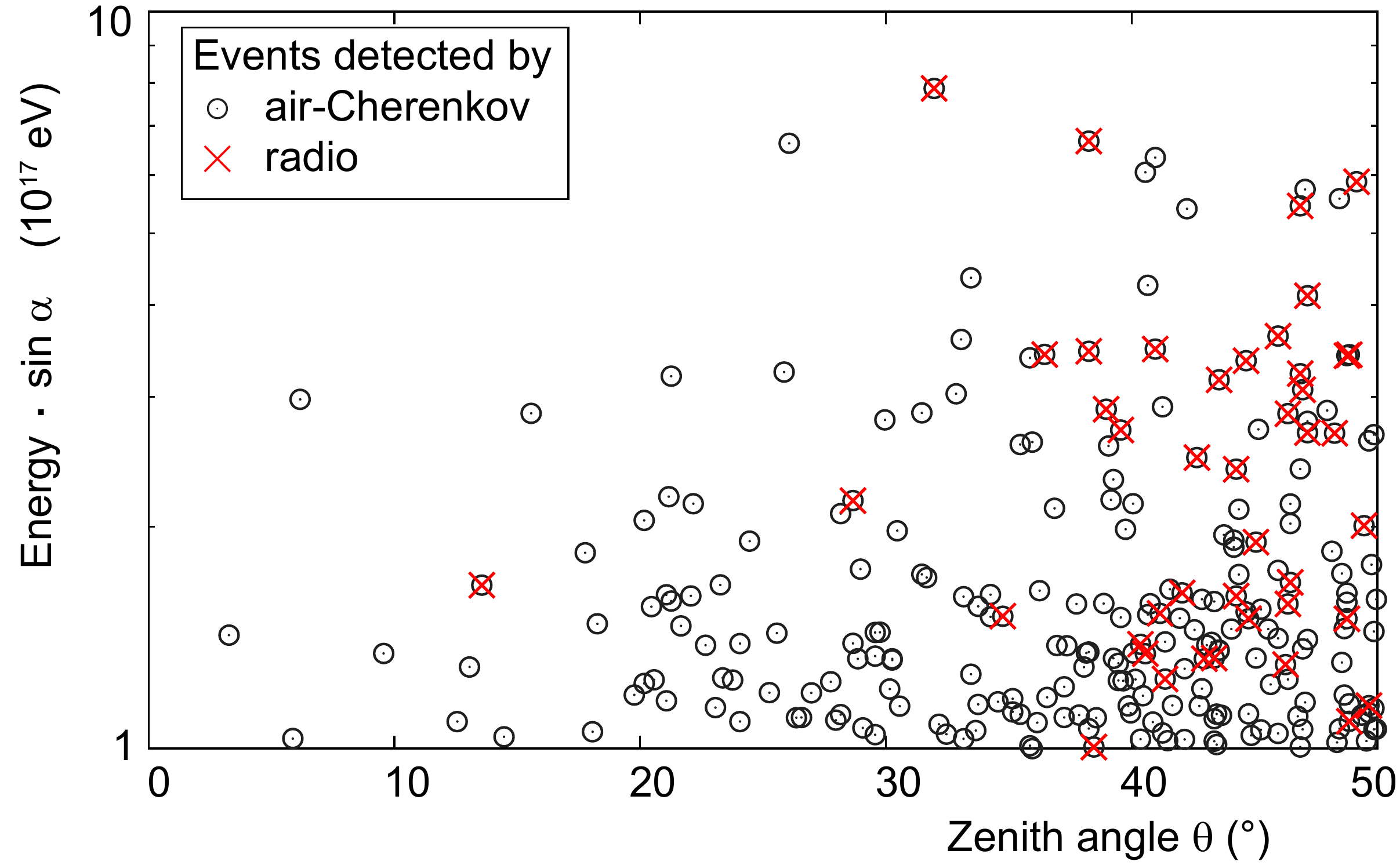}
\caption{Tunka events with energies above $10^{17}\,$eV detected by the PMT array since Tunka-Rex is running, and those events passing the quality cuts for the radio measurements. The efficiency increases with energy $E$, the sine of the geomagnetic angle $\sin \alpha$, and the zenith angle $\theta$.} \label{fig_efficiency}
\end{figure}

\begin{figure}[p!]
\centering
\includegraphics[width=\linewidth]{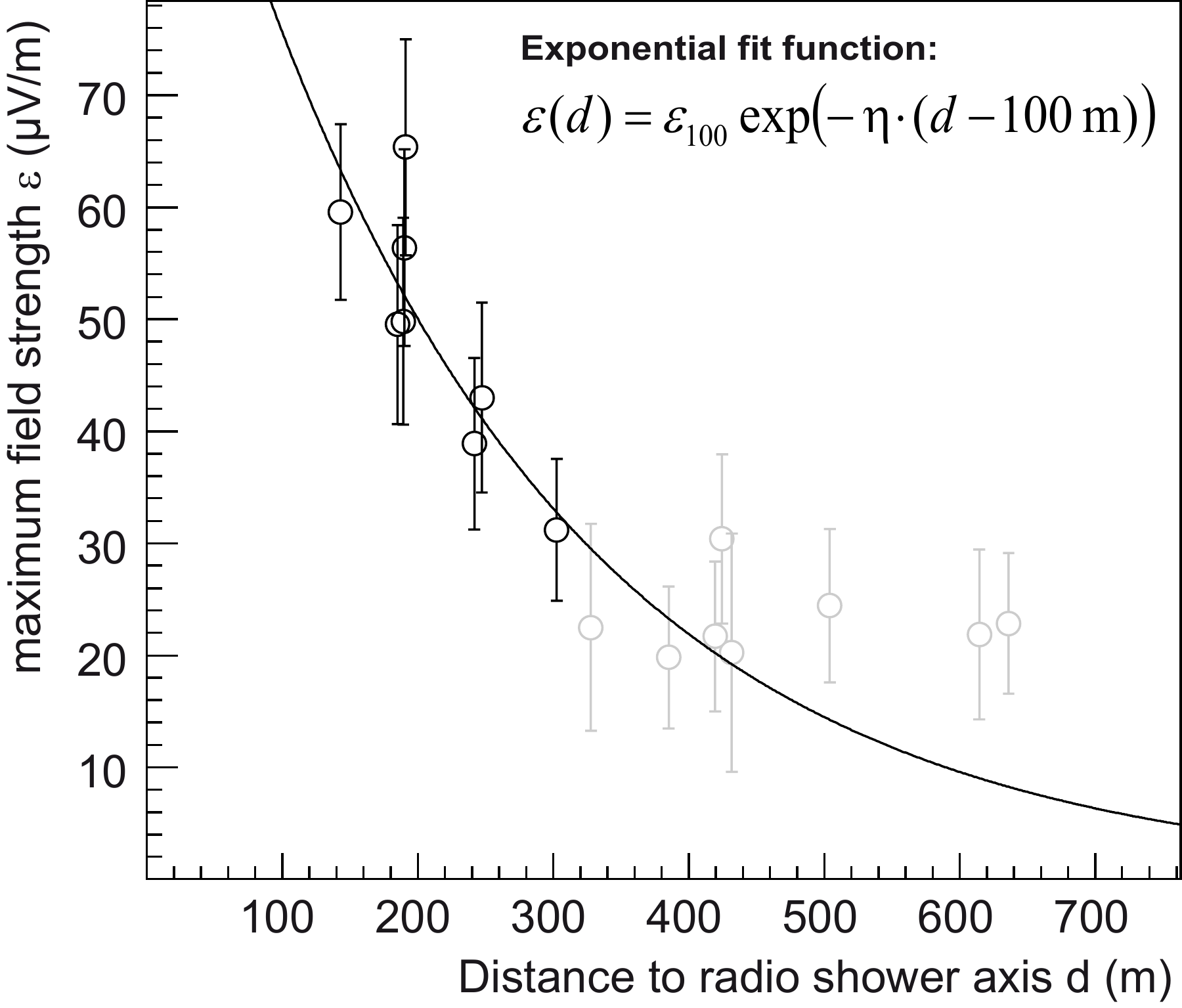}
\caption{Lateral distribution of the radio amplitude (= maximum field strength) for the example event shown in Fig.~\ref{fig_exampleEvent}.} \label{fig_exampleLDF}
\end{figure}

\section{First Results}
Until now, we found 49 events with a zenith angle $\theta \le 50^\circ$, and 82 events with $\theta > 50^\circ$ in an effective measurement time of 392 hours (Table \ref{tab_EventStatistics}). The distinction between more and less inclined showers is necessary, since the air-Cherenkov array can provide a good estimate for the energy and $X_\mathrm{max}$ only for the less inclined events, i.e.~only these events can be used for the cross-calibration of the radio and the air-Cherenkov measurements. Generally, the radio efficiency increases not only with large geomagnetic angles $\alpha$, i.e.~the angle between the shower axis and the geomagnetic field (Figs.~\ref{fig_angularDistribution} and \ref{fig_efficiency}), but also with larger zenith angles. Thus, once the cross-calibration is performed, we can use the inclined events to increase the total statistics of Tunka measurements at high energies. In addition, we expect that the event rate will increase when we complete the array to 25 antennas this autumn, and optimize our algorithms for digital background suppression.

To test the expected sensitivity of the Tunka-Rex measurements to air shower parameters, in particular to the energy, we reconstructed the lateral distribution of the radio signal for the 49 events with $\theta \le 50^\circ$. In a first approach, we used the shower geometry provided by the denser air-Cherenkov array to calculate the distance from each antenna to the shower axis, and then plotted the maximum absolute value of the electric field-strength vector as function of this axis distance. To estimate the uncertainties of the amplitude measurements and to correct the measured amplitudes for a bias due to background, we used formulas developed for the east-west aligned antennas of LOPES \cite{SchroederLOPESnoise_ARENA2010}, and then fitted an exponential function (Fig.~\ref{fig_exampleLDF}).

\begin{figure}
\centering
\includegraphics[width=\linewidth]{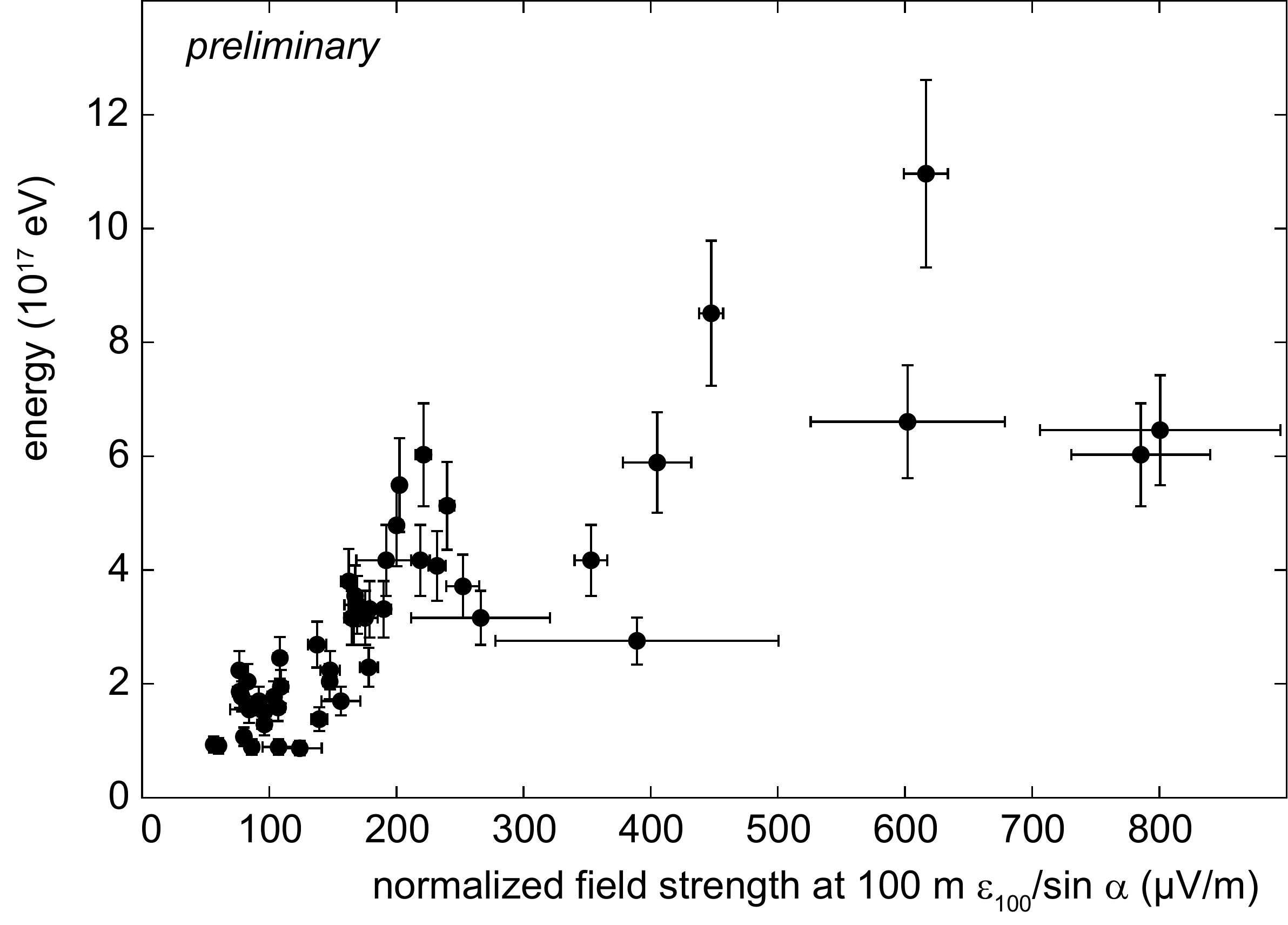}
\caption{Correlation between the radio field strength at 100 m corrected for the geomagnetic effect and the energy reconstructed with the air-Cherenkov measurements. The given values for the field strength are based on a preliminary calibration.} \label{fig_energy}
\end{figure}

Consistent with several historic and modern experiments \cite{Allan1971, SchroederLOPES_ARENA2012, GlaserAERA_ARENA2012, RebaiCODALEMAenergy2012}, the amplitude parameter of the lateral distribution is correlated with the primary energy (Fig.~\ref{fig_energy}). However, the analysis is still preliminary, e.g., because of the preliminary status of the calibration and because the impact of the background at Tunka has to be studied in more detail. Moreover, we expect that the slope of the lateral distribution is sensitive to the position of the shower maximum \cite{HuegeUlrichEngel2008, deVries2010}, which we will analyze in near future by comparing Tunka-Rex measurements to the $X_\mathrm{max}$ reconstruction of the PMT array Tunka-133. Future work will be dedicated to find an optimal reconstruction algorithm for the energy and $X_\mathrm{max}$, and to test the achievable precision by comparison to the air-Cherenkov measurements. In addition to using the lateral distribution, $X_\mathrm{max}$ might also be obtained form the radio measurements via the shape of the radio wavefront \cite{Lafebre2010, SchroederWavefrontICRC2011}, or the slope of the frequency spectrum \cite{GrebeAERA_ARENA2012}.

\section{Conclusion}
Tunka-Rex started operation in autumn 2012, and successfully measured the radio emission of air showers with energies above $10^{17}\,$eV in combination with the Tunka air-Cherenkov array. Thus, Tunka provides hybrid measurements which enable a cross-calibration between the air-Cherenkov and the radio signal. Our measurements are compatible with the picture that the radio emission originates mainly from the geomagnetic deflection of the electrons and positrons in the air shower \cite{KahnLerche1966}, since the efficiency depends on the energy and the arrival direction. Moreover, Tunka-Rex offers  ideal conditions to test the $X_\mathrm{max}$ sensitivity and precision by comparing the radio reconstruction to the air-Cherenkov measurements.

In future, we plan to optimize the reconstruction techniques, and to compare our measurements to simulations and other experiments. Moreover, we plan to trigger Tunka-Rex also by the scintillator extension of Tunka \cite{TunkaICRC2013}, and thus can measure also during day and increase the duty cycle by an order of magnitude. Finally, we collaborate with Tunka-HiSCORE \cite{HiSCORE_ICRC2013}, and will equip some of the HiSCORE detector stations in the Tunka array with additional antennas. By this we can study to which extent a denser array of radio detectors can increase the detection efficiency and the precision for the energy and for the mass composition.

\vspace*{0.5cm}
\footnotesize{{\bf Acknowledgment:}{This work was supported by the Russian Federation Ministry of Education and Science (G/C 14.518.11.7046, 14.B25.31.0010, 14.14.B37.21.0785, 14.B37.21.1294), the Russian Foundation for Basic Research (Grants 11-02-00409, 13-02-00214, 13-02-12095, 13-02-10001) and by the Helmholtz Association (Grant HRJRG-303) and the Deutsche Forschungsgemeinschaft (Grant TL51-3).}

\clearpage


\begin{thebibliography}{}
\bibitem{FalckeNature2005}H.~Falcke, \etal (LOPES Collaboration), Nature \textbf{435} (2005) 313.

\bibitem{SchroederLOPES_ARENA2012}F.~G.~Schr\"oder, \etal (LOPES Collaboration), Proc. 5th ARENA, Erlangen, Germany, AIP Conf. Proc. \textbf{1535} (2013) 78.

\bibitem{GlaserAERA_ARENA2012}C.~Glaser, for the Pierre Auger Collaboration, Proc. 5th ARENA, Erlangen, Germany, AIP Conf. Proc. \textbf{1535} (2013) 68.

\bibitem{RebaiCODALEMAenergy2012}A.~Rebai, \etal, arXiv.org (2012) 1210.1739.

\bibitem{ApelLOPES_MTD2012}W.~D.~Apel, \etal (LOPES Collaboration), Phys. Rev. D \textbf{85} (2012) 071101(R).

\bibitem{PalmieriLOPES_ICRC2013}N.~Palmieri, \etal (LOPES Collaboration), paper 0439, these proceedings.

\bibitem{TunkaICRC2013}N.~Budnev, for the Tunka Collaboration, paper 0418, these proceedings.

\bibitem{AERAantennaPaper2012}The Pierre Auger Collaboration, JINST \textbf{7} (2012) P10011.

\bibitem{HillerTunkaRex_ICRC2013}R.~Hiller, \etal (Tunka-Rex Collaboration), paper 1278, these proceedings.

\bibitem{AugerOffline2007}S.~Argiro, \etal, Nucl. Instr. Meth. A \textbf{580} (2007) 1485.

\bibitem{RadioOffline2011}The Pierre Auger Collaboration, Nucl. Instr. Meth. A \textbf{635} (2011) 92.

\bibitem{SchroederLOPESnoise_ARENA2010}F.~G.~Schr\"oder, \etal (LOPES Collaboration), Nucl. Instr. Meth. A \textbf{662} (2012) S238.

\bibitem{Allan1971}H.~R.~Allan, Prog. in Elem. Part. and Cosmic Ray Phys. \textbf{10} (1971) 171.

\bibitem{HuegeUlrichEngel2008}T.~Huege, R.~Ulrich, R.~Engel, Astropart. Phys. \textbf{30} (2008) 96.

\bibitem{deVries2010}K.~D.~de Vries, \etal, Astropart. Phys. \textbf{34} (2010) 267.

\bibitem{Lafebre2010}S. Lafebre, \etal, Astropart. Phys. Astropp. \textbf{34} (2010) 12.

\bibitem{SchroederWavefrontICRC2011}F.~G.~Schr\"oder, \etal (LOPES Collaboration), Proc. 32nd ICRC Beijing, China (2011) \#0313.

\bibitem{GrebeAERA_ARENA2012}S.~Grebe, for the Pierre Auger Collaboration, Proc. 5th ARENA, Erlangen, Germany, AIP Conf. Proc. \textbf{1535} (2013) 73.

\bibitem{KahnLerche1966}F.~D.~Kahn, I.~Lerche, Proc. Roy. Soc. A \textbf{289} (1966) 206.

\bibitem{HiSCORE_ICRC2013}R.~Wischnewski, \etal (HiSCORE Collaboration), paper 1164, these proceedings.
\end{thebibliography}
\end{document}